\documentclass{optica-article}

\journal{opticajournal} 

\articletype{Research Article}

\usepackage{siunitx}
\begin{document}

\title{A Compact, Robust, and Tunable Open Microcavity Platform for Solid-State Quantum Electrodynamics}

\author{Thi D. Hoang\authormark{1}, Fateme Mahdikhany\authormark{2}, Zixuan Wang\authormark{3}, Richard Mirin\authormark{3}, Kevin Silverman\authormark{3}, Poolad Imany\authormark{2}, and Shuo Sun\authormark{1,*}}

\address{\authormark{1}JILA and Department of Physics, University of Colorado Boulder, Boulder CO 80309, USA\\
\authormark{2}Icarus Quantum Inc., Boulder 80302, USA\\
\authormark{3}National Institute of Standards and Technology, Boulder CO 80305, USA}

\email{\authormark{*}shuosun@colorado.edu} 

\begin{abstract*} 
Open microcavities provide a powerful platform for studying cavity quantum electrodynamics in solid-state systems. However, operating open microcavities at cryogenic temperatures, as required for many solid-state quantum emitters, typically demands bulky and cryostat-specific vibration-mitigation setups. Here we report a compact, robust, and tunable mechanical host for an open microcavity. The complete mechanical assembly fits within a footprint of $1'' \times 1'' \times 0.5''$. Using this mechanical host, we observe no vibration-induced cavity broadening for an open microcavity with finesse exceeding 1,000 without cryostat customization or active locking. The assembly also enables in situ tuning of the cavity resonance over 3 nm, and the resonance of the same cavity remains within this range across multiple cooldowns. To further showcase the capability of this assembly, we demonstrate coupling between an InGaAs quantum dot and an open microcavity with a cooperativity exceeding unity. This platform provides a versatile testbed for fundamental cavity quantum electrodynamics and a scalable route to portable quantum light sources and spin–photon interfaces for quantum repeaters, quantum networks, and photonic quantum computing.
\end{abstract*}

\section{Introduction}
The open microcavity, a Fabry–Perot cavity formed between a flat mirror and a concave micromirror, is a powerful platform for studying cavity quantum electrodynamics (QED) in solid-state systems~\cite{li2019tunable}. Its design combines a high quality factor with a small mode volume, both of which are essential for strong light–matter interactions~\cite{janitz2020cavity}. In contrast to most nanophotonic cavities, an open microcavity requires no etching near the emitter, thereby preserving emitter coherence and spectral stability. The cavity length can be tuned in situ to match the cavity resonance with the emitter’s optical transition, which is especially valuable for solid-state emitters with inhomogeneous emission spectra. The fundamental cavity mode naturally exhibits high spatial overlap with the Gaussian mode and thus couples efficiently to single-mode fibers. Owing to its open architecture, open microcavities are compatible with a broad range of solid-state emitters and materials and have been used to integrate quantum dots~\cite{muller2009coupling,miguel-sanchez2012cavity, najer2019gated, tomm2021bright,tomm2024realization, maisch2024investigation,ding2025highefficiency}, color centers in diamond~\cite{benedikter2017cavityenhanced,dolan2018robust, bayer2023optical,herrmann2024coherent, pallmann2024cavitymediated, yurgens2024cavityassisted, zifkin2024lifetime, fischer2025spinphoton} and silicon carbide~\cite{hessenauer2025cavity}, rare-earth ions~\cite{casabone2018cavityenhanced, merkel2020coherent, casabone2021dynamic, gupta2025dual}, 2D materials~\cite{vogl2019compact, vadia2021opencavity,drawer2023monolayerbased}, and organic molecules~\cite{wang2019turning}. When coupled to a quantum dot, open microcavities have enabled the deepest strong coupling among optical cavity QED systems~\cite{najer2019gated} and the highest source-to-detector efficiency for a deterministic single-photon source~\cite{tomm2021bright,ding2025highefficiency}.

Despite these advantages, it remains challenging to operate open microcavities at cryogenic temperatures, as required for many solid-state quantum emitters. The cavity performance is extremely sensitive to mechanical vibration: even for a cavity finesse of $10^3$, a 1-nm change in cavity length shifts the resonance by one linewidth for a cavity mode at wavelength $\lambda_c = 1~\mu\text{m}$. One way to mitigate this sensitivity is to use monolithic Fabry–Perot microcavities~\cite{engel2023purcell, yang2024tunable}, which rigidly define the cavity length and thus offer excellent mechanical robustness. However, monolithic cavities eliminate in-situ tuning of the cavity resonance, making it difficult to match the cavity to solid-state quantum emitters with inhomogeneous optical transitions. To retain cavity tunability, many experiments instead employ liquid-helium bath (open-cycle) cryostats~\cite{muller2009coupling, miguel-sanchez2012cavity, najer2019gated, wang2019turning, tomm2021bright,bayer2023optical, maisch2024investigation, yurgens2024cavityassisted}, yet these systems are increasingly uneconomical and difficult to scale. Using a standard closed-cycle cryostat demands substantial vibration-isolation engineering for the mechanical assembly that hosts the tunable open microcavity~\cite{merkel2020coherent,fontana2021mechanically,vadia2021opencavity,ruelle2022tunable,pallmann2023highly, herrmann2024coherent}, sometimes in combination with active locking~\cite{wang2019turning, merkel2020coherent, casabone2021dynamic, vadia2021opencavity}, resulting in bulky setups and additional lasers. These constraints hinder the scalable and practical deployment of open-microcavity-based cavity QED systems for optical quantum computing~\cite{duan2004scalable,kok2007linear,o2007optical}, quantum networking~\cite{reiserer2022colloquium,azuma2023quantum}, and photonic quantum metrology~\cite{pirandola2018advances,couteau2023applications}. Moreover, the bulky mechanical assemblies used to host tunable open microcavities does not fit inside the bore of a standard superconducting magnet ($\sim 1 - 2''$ in diameter), making the study of spin–photon interfaces for emitters such as III-V quantum dots impractical~\cite{sun2016quantum,luo2019spin}.

In this paper, we demonstrate a compact, robust, and tunable open microcavity platform for scalable operation in standard closed-cycle cryostats. The complete mechanical assembly fits within a $1'' \times 1'' \times 0.5''$ footprint and can be mounted in virtually any cryostat, including inside the bore of a superconducting magnet. Using this assembly to host an open microcavity, we observe no vibration-induced cavity broadening for a cavity with finesse exceeding 1,000, even in a standard closed-cycle cryostat without vibration isolation or active locking. The cavity resonance can be tuned in situ over about 3 nm in wavelength, and across six thermal cycles the resonance wavelength of a given cavity remains within this tuning range, showing that the assembly can be pre-characterized, packaged, and reliably deployed in different systems. To further showcase the capabilities of this platform, we demonstrate coupling between the open microcavity and an InGaAs quantum dot, a promising quantum emitter with near-unity quantum efficiency, large dipole moments, and excellent spectral stability~\cite{senellart2017high}. We routinely observe cooperativities close to or larger than unity, limited only by the cavity design rather than the mechanical assembly. This platform thus provides a versatile testbed for cavity QED and a scalable route to portable quantum light sources~\cite{aharonovich2016solid} and spin–photon interfaces~\cite{lodahl2015interfacing,awschalom2018quantum} for photonic quantum computers~\cite{duan2004scalable,kok2007linear,o2007optical}, quantum repeaters~\cite{azuma2023quantum}, and quantum networks~\cite{reiserer2022colloquium}.

\section{Design of the cavity assembly}

\begin{figure}
    \centering
    \includegraphics[width=\linewidth]{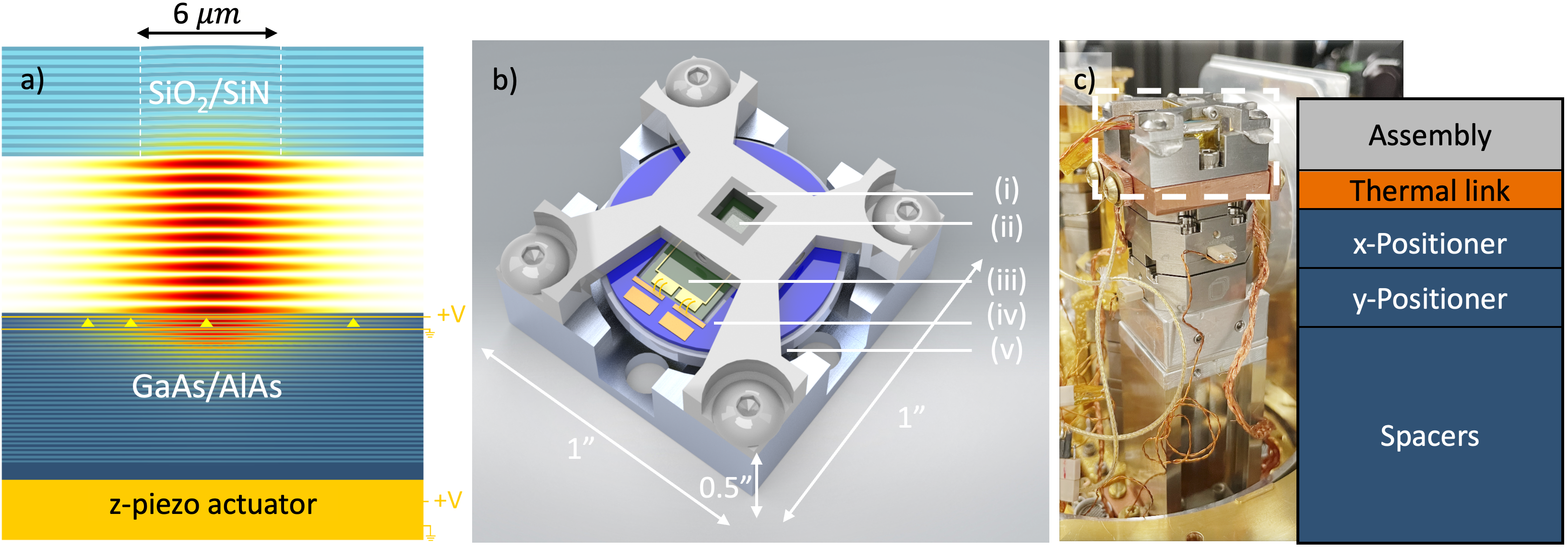}
    \caption{(a) Schematic of the open microcavity used in our work, overlaid with the calculated electric field profile of its fundamental mode based on the cavity length estimated from experiment. The cavity is formed by a planar DBR with 29.5 pairs of alternating GaAs/AlAs layers and a microscale concave DBR with 12.5 pairs of alternating $\mathrm{SiO}_2$/SiN layers. A GaAs epitaxial layer with InGaAs quantum dots (yellow triangles, enlarged for visibility) embedded in a p-i-n diode (yellow solid lines, enlarged for visibility) is grown on top of the planar DBR. The lateral extent of the fabricated concave feature, marked by the white dashed lines, is approximately 6~$\mu$m. (b) Design of the mechanical assembly used to host the open microcavity. The assembly consists of a top and a bottom holder, fastened together by four corner screws. The top mirror substrate (ii) is glued onto a metal square aperture (i), which is attached to the top holder with set screws. The bottom chip (iii) sits on a ring-shaped piezo (v) and inside a printed circuit board (purple, iv). (c) Photograph of the open microcavity assembly mounted inside a commercial closed-cycle cryostat.}
    \label{fig:fig1}
\end{figure}

Figure~\ref{fig:fig1}(a) shows the schematic structure of the open microcavity used in our work. The cavity is a microscale plano–concave Fabry–Perot resonator. The planar mirror on the bottom is a distributed Bragg reflector (DBR) consisting of 30 pairs of alternating GaAs/AlAs layers grown by molecular beam epitaxy (MBE). A p–i–n junction formed by p-doped, intrinsic, and n-doped GaAs with InGaAs quantum dots embedded in the intrinsic layer is grown on top of the planar DBR (see Supplementary Section 1). The p–i–n structure enables DC Stark shift of the quantum dot resonance via an applied bias voltage, and this capability is preserved for quantum dots coupled with our open microcavities (see Supplementary Section 2). The top mirror is fabricated by first using the resist-reflow technique~\cite{jin2022microfabricated} to define concave profiles with a minimum radius of curvature of $50~\mu$m, and then depositing 12.5 pairs of $\mathrm{SiO}_2$/SiN by plasma-enhanced chemical vapor deposition (PECVD) (see Supplementary Section 3). Atomic force microscopy (AFM) measurements show an rms surface roughness of 0.5 nm after PECVD coating, limited by our AFM resolution (see Supplementary Section 4). We note that PECVD coating will likely lead to a worse surface roughness compared with ion beam sputtering~\cite{jin2022microfabricated}. Nevertheless, we found PECVD to be more accessible and faster than ion beam sputtering while still routinely delivering cavity finesse exceeding $10^3$, which is sufficient to reach the high-cooperativity regime discussed in Sec.~\ref{sec:sec4}. The overlay in Fig.~\ref{fig:fig1}(a) shows the calculated electric field profile of the fundamental cavity mode, which predicts a quality factor of $Q = 6.0\times 10^4$, a mode volume of $11\lambda_0^3$ where $\lambda_0 = 971$~nm is the cavity resonant wavelength in vacuum, and a finesse of 5,500 for the cavity length used in our experiment. The cavity is designed to be one-sided, where the reflectivity of the bottom mirror (99.99\%) is much larger than that of the top mirror (99.93\%).

Figure~\ref{fig:fig1}(b) shows the design of the mechanical assembly used to host the open microcavity. Although the design is optimized for the specific cavity used in this work, the assembly is general and can be adapted to support a wide range of open microcavity geometries and quantum emitters. The mechanical structure consists of two principal parts held together at the four corners, hereafter referred to as the top holder and the bottom holder. The top mirror substrate (ii) is glued onto a square metal aperture (i) with the same outer dimensions, and this aperture is clamped to the top holder with set screws on its four sides. At the bottom, the GaAs chip (iii) sits in the center of a printed circuit board (iv). Wire bonding between (iii) and (iv) provides electrical connection to apply an electric field across the p-i-n junction. Both components are glued onto a commercially available ring-shaped piezo (v) (Thorlabs PA44M3KW) that is attached to the bottom holder. Finally, the top holder is fastened to the bottom holder with four corner screws. Once secured, the mechanical assembly does not allow translational motion between the two mirror substrates. However, we can tune the cavity length in situ using the ring-shaped piezo, which moves only the bottom substrate along the optical axis. 

Several design considerations are implemented to ensure high robustness against vibration. The assembly is largely symmetric to minimize thermal distortion and maximize mechanical stability, with the only asymmetry being a cutout in the top holder that provides access for electrical connections to the chip. All components are either glued or rigidly clamped with screws, which is critical for stability and reproducibility during transportation and thermal cycling. The two mirror substrates are aligned parallel and brought into contact by monitoring interference fringes (Newton rings, see more details in Supplementary Section~5). Currently, all metal parts are made of stainless steel to provide high rigidity and low thermal contraction. They can be readily replaced with titanium if operation in large magnetic fields is required.

Figure~\ref{fig:fig1}(c) shows a photo of the mechanical assembly mounted inside a standard closed-cycle cryostat (Montana Instruments s100). The assembly occupies a footprint of only $1'' \times 1'' \times 0.5''$ with a mass of 34.6 grams. The compactness of the structure makes it feasible to mount the cavity in virtually any commercial cryostat with constrained sample space, including inside the bore of a superconducting magnet. In addition, the mass of the assembly is well within the load limit of standard commercial cryogenic nanopositioners. In our experiment, we mount the cavity assembly on two nanopositioners (Attocube ANPx101) that translate the cavity in the transverse plane relative to the optical axis. The distance between the cavity and the objective lens is controlled by moving the objective lens outside the cryostat, which is not shown in the figure.

\section{Performance of the cavity assembly}
We now characterize the performance of the cavity assembly at cryogenic temperature. We first calibrate the cavity length, defined as the number of wavelengths spanned between the two DBR mirrors. For open microcavities with very short cavity length, it is generally difficult to determine this quantity accurately, because the free spectral range (FSR) typically used to infer the cavity length is comparable to or larger than the DBR bandgap. When the open microcavity is operated near the center of the DBR bandgap, as is desirable for maximizing the cavity $Q$, the adjacent longitudinal mode lies outside the DBR bandgap, making the FSR inaccessible. In our case, we take advantage of the large array of concave mirrors on the top substrate and the small variation in cavity length across different cavities on the same chip. We identify one cavity in which two longitudinal modes sit near the two edges of the DBR bandgap, measure its FSR, and use the inferred cavity length as a rough estimate for neighboring cavities on the same chip. We emphasize that this cavity length is only an estimate, but it should be accurate to within approximately half a wavelength.

\begin{figure}[h!]
    \centering
    \includegraphics[width=1\linewidth]{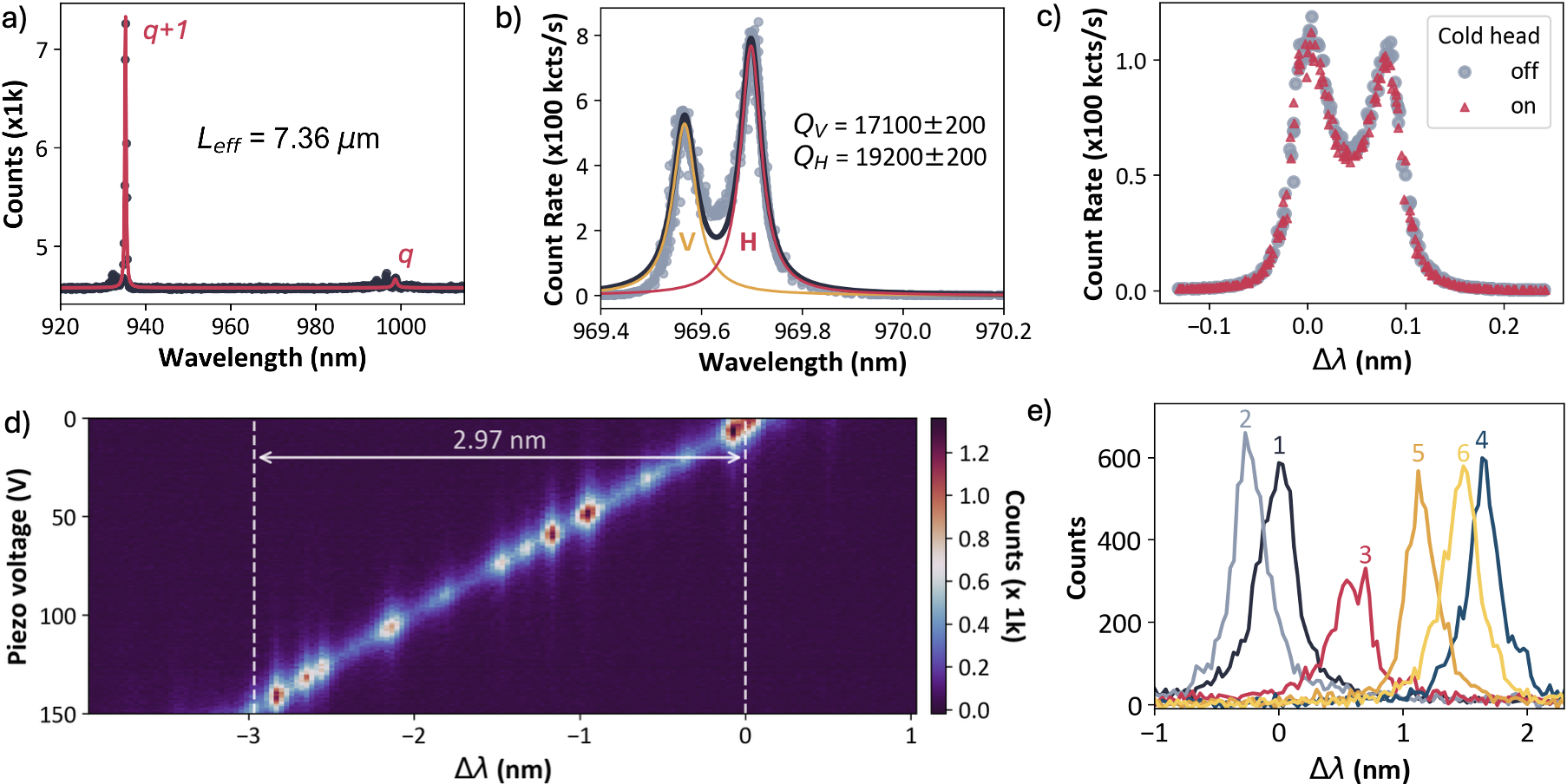}
    \caption{Characterization of the open microcavity and its mechanical host. (a) Spectrum of a cavity with two adjacent longitudinal modes lying within the DBR bandgap, obtained via photoluminescence spectroscopy of the quantum dot ensemble. (b) Cross-polarized transmission spectrum of a different cavity at the center of the DBR bandgap, showing two non-degenerate polarization modes. (c) Transmission spectrum of the same cavity when the compressor of the cryostat is on (red triangles) and off (grey circles). The two spectra are nearly identical, showing that we have no vibration induced cavity broadening. (d) Photoluminescence spectrum of the cavity as we vary the bias voltage applied on the piezo. We observe a continuous cavity resonance shift with a tuning range of 3.0 nm. (e) Photoluminescence spectrum of the same cavity mode across 6 different thermal cycles. The count fluctuations among different cooldowns are due to changes in optical collection efficiency, as the assembly can shift slightly in position and angle after repeated thermal cycling.}
    \label{fig:fig2}
\end{figure}

Figure~\ref{fig:fig2}(a) shows the spectrum of a cavity whose full FSR lies within the DBR bandgap. We obtain this spectrum by exciting the cavity with a 780-nm laser and collecting the photoluminescence from the embedded quantum dots. Owing to the high density and large inhomogeneous linewidth of the quantum dots, the quantum dot ensemble acts as a broadband internal light source that illuminate the cavity and reveal the cavity modes. Fitted in red are the fundamental transverse modes with consecutive longitudinal mode numbers, $q+1$ and $q$. We also observe weaker satellite peaks on the blue side of each fundamental mode, corresponding to higher-order transverse cavity modes that are collected into the signal fiber due to imperfect mode matching between the cavity fundamental mode and the collection fiber. From the measured spectrum, we extract an FSR of 63 nm, defined as the measured difference in wavelength between two adjacent longitudinal modes. When two adjacent longitudinal modes are near the edges of the DBR bandgap, as is the case for our cavity, the FSR does not simply relate to the cavity length via $\Delta\nu_{\text{FSR}} = c/(2nL)$, or equivalently $\Delta\lambda_{\text{FSR}}=\lambda^2/(2nL)$, due to the frequency-dependent phase shift induced by each mirror near the band edge (see Supplementary Section 6)~\cite{koks2021microcavity}. Nevertheless, this measurement shows an upper bound of the cavity length to be $(nL)_{\text{max}}=\lambda_{q+1}\lambda_q/(2\Delta\lambda_{\text{FSR}})=7.36~\mu \text{m}$, with $\lambda_{q+1}, \lambda_q$ being the resonant wavelengths of two adjacent longitudinal modes, and $\Delta\lambda_{\text{FSR}}=\lambda_q-\lambda_{q+1}$. The cavity length value includes the length of the airgap, material in between the two DBRs, and penetration depth into the DBRs. 

We next measure the cavity $Q$ factor. Here we select a different cavity whose resonance lies near the center of the DBR bandgap. We excite the cavity with a tunable, narrow-linewidth continuous-wave laser and collect the reflected signal using a cross-polarization detection scheme. Figure~\ref{fig:fig2}(b) shows the resulting cavity response as a function of laser frequency. We observe two cavity resonances, corresponding to orthogonally polarized modes whose degeneracy is lifted by the anisotropic refractive index of GaAs, consistent with previous reports on similar cavities~\cite{tomm2021bright, ding2025highefficiency}. The extracted quality factors of the two modes are $Q_V=\num{17100}\pm200$ and $Q_H=\num{19200}\pm200$, approximately one third of the simulated values. This reduction in $Q$ comes from a few factors: the surface roughness of the concave mirror, the mirror's reflectivity stopband being off from the resonant wavelengths, and an overestimate of the cavity length in the simulation. From these measurements, we estimate a lower bound of the cavity finesses to be 1,100 and 1,200 for the two modes.

Figure~\ref{fig:fig2}(c) shows the spectrum of the cavity measured with the cryostat compressor on (red triangles) and off (gray circles). The two spectra are nearly identical, with no noticeable cavity broadening when the compressor is on. This demonstrates that, at a finesse exceeding 1,000, our cavity assembly is sufficiently robust against mechanical vibrations in a standard closed-cycle cryostat without requiring cryostat modifications for passive vibration mitigation or active cavity locking.

An important feature of our cavity design is its ability to tune the resonance in situ. To characterize the tuning range, we measure the cavity spectrum via photoluminescence spectroscopy of the quantum dot ensemble while varying the voltage applied to the piezo inside the cavity assembly. Figure~\ref{fig:fig2}(d) shows the measured cavity spectra as a function of piezo voltage. We observe a continuous shift of the cavity resonance over 3.0~nm at 4~K. The cavity peak intensity varies strongly across the tuning range, reflecting variations in the coupling strength between the cavity mode and quantum dots across this wavelength range. This arises because quantum dots emitting at different wavelengths occupy different positions relative to the cavity mode. The tuning range can be further extended by applying a higher voltage to the piezo or by using a longer piezo stack. In Supplementary Section~7, we show a similar measurement for a modified cavity assembly incorporating a longer piezo, which achieves a tuning range of 7~nm.

Currently, the in situ tuning range of our cavity (a few nanometers) is small compared with the inhomogeneous linewidth of the quantum-dot ensemble (tens of nanometers). This limited tuning range, together with the lack of independent translational motion between the top and bottom substrates, makes deterministic coupling to an arbitrary quantum dot difficult. Nevertheless, with an optimized quantum dot density, the probability of finding a dot whose position lies within the Gaussian waist of the cavity mode and whose emission falls within the tuning range is reasonably high. In that case, taking advantage of the large array of cavities on the same chip, one could pre-characterize many devices, identify and label those that perform well (for example, as single-photon sources), and then transport the entire assembly to a user facility. For this strategy to be practical, the cavity resonance must remain stable to within the tuning range over repeated thermal cycles. To explore this possibility, Fig.~\ref{fig:fig2}(e) shows the photoluminescence spectra of the same cavity across six successive thermal cycles, with the piezo actuator held at a fixed voltage. The cavity mode remains within a 2.5-nm wavelength window, indicating that the cavity length is robust against thermal cycling. Since this variation is comparable to our in situ tuning range, a pre-characterized device can be reliably relocated and used in a different facility. 

We note that, in contrast to the cavity transmission spectra shown in Figs.~\ref{fig:fig2}(b) and \ref{fig:fig2}(c), which were obtained by laser scanning, the cavity photoluminescence spectra shown in Figs.~\ref{fig:fig2}(a), \ref{fig:fig2}(d), and \ref{fig:fig2}(e) do not show the doublet feature arising from the two non-degenerate polarization modes. This is because the spectrometer resolution used for the photoluminescence measurements was insufficient to resolve the doublet. In Supplementary Section~8, we show that the cavity doublet can be resolved under photoluminescence spectroscopy when a sufficiently high spectrometer resolution is used.

\section{Cavity QED experiments}
\label{sec:sec4}
\begin{figure}[!h]
    \centering
    \includegraphics[width=0.85\linewidth]{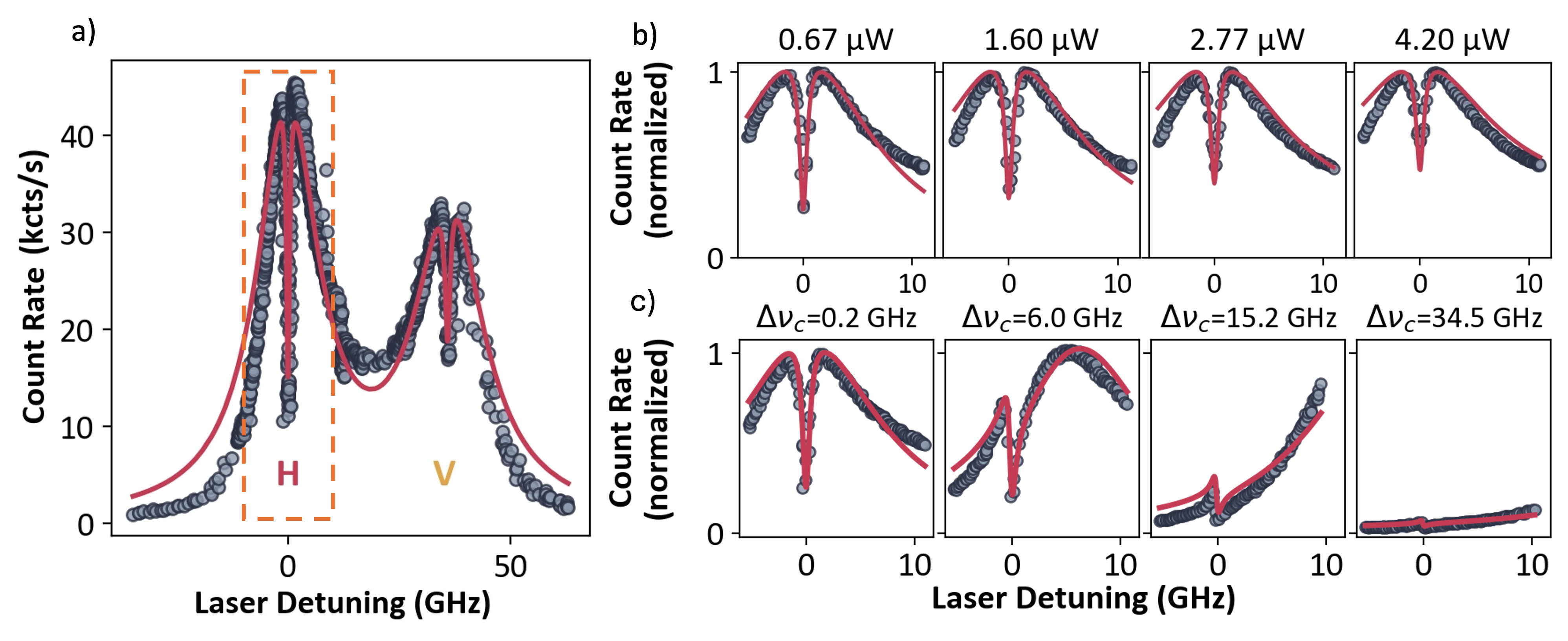}
    \caption{Measurement of a cavity QED device with single quantum dots coupled to an open microcavity. (a) Cross-polarized cavity transmission spectrum of an open microcavity coupled with two distinct quantum dots. (b) Cross-polarized transmission spectrum of the $H$-polarized cavity mode (corresponding to the the dashed-line box region of (a)) as we increase the power of the incident laser. The power is measured right before the objective lens. (c) Cross-polarized transmission spectrum of the $H$-polarized cavity mode as we vary the detuning between the cavity and the quantum dot by varying the cavity length. In all panels, grey circles are measured data, and red solid lines are the theoretical fit.}
    \label{fig:fig3}
\end{figure}

We now employ our cavity assembly to demonstrate controlled coupling between single InGaAs quantum dots and open microcavities. We first identify a cavity that couples to a quantum dot inside the mode volume. Figure~\ref{fig:fig3}(a) shows the cross-polarized transmission spectrum of this cavity, measured using the same tunable continuous-wave laser as in Fig.~\ref{fig:fig2}(b). The spectrum exhibits two broad peaks corresponding to the two non-degenerate polarization modes, but now each peak contains a narrow dip, characteristic of dipole-induced transparency when the cavity is coupled to a single quantum emitter~\cite{waks2006dipole}. In this device, the two dips arise from two distinct quantum dots, each resonant with one of the cavity modes. By fitting the measured spectrum (gray circles) to a theoretical model (red solid line), we extract a coupling strength of $g_H/2\pi = 1.37\pm 0.02$~GHz and a cooperativity of $C_H = 1.8\pm0.2$ for the $H$ mode (see Supplementary Section~9 for details of the theoretical fit). We have performed similar fits on several other devices, the extracted parameters of which are summarized in Table~1 and discussed in Supplementary Section~10. The highest cooperativity observed thus far is $2.8\pm0.5$.

\begin{table}[h]
    \centering
    \begin{tabular}{|c|c|c|c|c|c|}
        \hline
            Device \#
            & $\kappa/2\pi~\mathrm{(GHz)}$ & $g/2\pi~\mathrm{(GHz)}$ &  $\Gamma/2\pi~\mathrm{(GHz)}$ & $C=\frac{4g^2}{\kappa\Gamma}$\\ \hline 
        1 & $16.0 \pm 0.1$ &  $1.37 \pm 0.02$ &  $0.26\pm 0.08$ &  $1.8 \pm 0.2$ \\
        2 &  $13.5 \pm 0.5$ &  $1.83 \pm 0.07$ &  $0.4 \pm 0.1$ &  $2.5 \pm 0.6$\\
        3 &  $28.7\pm 0.7$ &  $1.82\pm 0.07$ &  $0.2\pm 0.3$ &  $2.8 \pm 0.5$ \\
        4 &  $13.6 \pm 0.9$ &  $1.0\pm 0.1$ &  $0.60 \pm 0.06$ &  $0.5\pm 0.1$ \\
        \hline
    \end{tabular}
    \caption{Summary of the cavity QED parameters for four different cavities on the same assembly. Device \#1 is the same as in Fig.~\ref{fig:fig3}. The raw sectra used to extract the cavity QED parameters for devices \#2 -- \#4 are provided in Supplementary Section~10.}
    \label{tab:cooperativity}
\end{table}

Figure~\ref{fig:fig3}(b) shows the spectrum of the $H$-polarized cavity mode as we increase the incident laser power. We observe clear saturation of the quantum-dot-induced dip, demonstrating the nonlinear quantum response of the cavity QED system. Our cavity assembly also allows us to control the detuning between the quantum dot and the cavity in situ. Figure~\ref{fig:fig3}(c) shows the spectra of the $H$-polarized cavity mode as we shift the cavity resonance. As the cavity is detuned from the quantum dot, the lineshape evolves from a high-contrast dip into a Fano-like feature. The measured data (gray circles) agree reasonably well with a theoretical model (red solid line) in which all parameters are fixed to the values obtained from Fig.~\ref{fig:fig3}(a), except for the cavity resonance frequency. 

With our relatively high quantum dot density, it is straightforward to identify cavities that couple to quantum dots with cooperativities close to or exceeding unity. In Supplementary Section 11, we showed the measured cavity spectra of 10 randomly selected cavities on the same mirror chip, each with a radius of curvature below 300~$\mu$m. We found that 8 out of 10 cavities exhibited a quantum-dot-induced transmission dip, with contrasts ranging from 25\% to 80\%.

\section{Discussion}

We demonstrated a compact, robust, and tunable mechanical assembly for operating open microcavities at cryogenic temperatures. With a footprint of only $1'' \times 1'' \times 0.5''$, in-situ resonance tuning over 3~nm, no measurable vibration-induced cavity broadening, and stable cavity resonances across multiple thermal cycles, our platform offers greater accessibility for studying solid-state cavity QED at cryogenic temperature. We showcase this capability by demonstrating the coupling between an open microcavity and an InGaAs quantum dot with a cooperativity exceeding unity. In addition, the compact footprint makes the system readily compatible with the bore of a superconducting magnet, enabling the studies of solid-state spin–photon interactions. Beyond these performance metrics, a key practical advantage of our platform is that the assembly can be easily transported between a development lab and a user facility, facilitating scalable deployment of quantum light sources~\cite{aharonovich2016solid} and spin–photon interfaces~\cite{lodahl2015interfacing,awschalom2018quantum} for photonic quantum computers~\cite{duan2004scalable,kok2007linear,o2007optical}, quantum repeaters~\cite{azuma2023quantum}, and quantum networks~\cite{reiserer2022colloquium}.

The cavity metrics demonstrated in this work, including the finesse, quality factor, mode volume, and cooperativity, are limited by the cavity itself rather than the host assembly. By depositing the concave DBR with ion beam sputtering~\cite{jin2022microfabricated}, we will improve the surface smoothness and increase the cavity $Q$ toward the design-limited value. Increasing the number of DBR pairs on each mirror would further boost the cavity $Q$. In parallel, by employing CO$_2$ laser ablation~\cite{muller2009coupling, hunger2010fiber, muller2010ultrahighfinesse, miguel-sanchez2012cavity, pallmann2024cavitymediated}, or focused-ion-beam milling~\cite{trichet2015topographic, vogl2019compact, wang2019turning, drawer2023monolayerbased} for fabrication of the concave mirror, we can make mirrors with a smaller radius of curvature and thereby reduce the cavity mode waist. The cavity length can also be reduced by decreasing the air-gap size, which can be achieved by defining the concave mirrors on a small mesa, so that a slight tilt between the two mirror substrates will not lead to a large air gap at the targeted cavity due to first contact occurring far away. With these improvements, we expect to achieve a finesse exceeding $10^5$ and cooperativity well above 100, although it will be important to investigate whether vibration-induced broadening remains negligible at such high finesse, which we leave for future work.

Beyond semiconductor quantum dots, our mechanical assembly can be easily adapted to form high-quality cavities around other solid-state emitters, such as defects in 2D materials~\cite{vogl2019compact, vadia2021opencavity,drawer2023monolayerbased}, color centers in diamond thin films~\cite{janitz2020cavity,herrmann2024coherent, pallmann2024cavitymediated, yurgens2024cavityassisted, zifkin2024lifetime, fischer2025spinphoton},  and organic molecules~\cite{wang2019turning}. The assembly can also be employed to couple open microcavities with other planar devices defined on a semiconductor chip, such as surface acoustic wave cavities~\cite{imany2022quantum,decrescent2024coherent} or microwave cavities~\cite{rochman2023microwave,xie2025scalable}, enabling new optomechanical systems and microwave–optical transducers mediated by quantum emitters. Currently, our mechanical assembly does not allow in situ translational alignment of the cavity to the quantum emitter, which may present challenges for scalability in some applications. This challenge can be partially addressed by either developing an automated method to pre-characterize a large array of devices and work with pre-selected cavities, or by aligning the cavity with the emitter at room temperature before securing the assembly for cooldown. The latter approach would require in-situ room-temperature scanning photoluminescence imaging during cavity assembly process, deterministic emitter placement or fabrication~\cite{sipahigil2016integrated, bayer2023optical, ngan2023quantum}, or alignment markers that label the emitter positions~\cite{sapienza2015nanoscale}. In addition, for translationally invariant material systems, such as excitons in quantum wells~\cite{gibbs2011exciton}, the lack of translational alignment is not a limitation. Taken together, we expect our platform to enable scalable and deployable implementations of cavity QED and hybrid quantum transduction across a wide range of solid-state emitters.

\section*{Funding}
We acknowledge funding from the National Science Foundation (2137953 and 2016244), the Air Force Office of Scientific Research (AFOSR) (FA2386-24-1-4067), NASA STTR (80NSSC23PB442), and Colorado Quantum Seed Grant (1564972). S.S. acknowledges support from the NSF CAREER Award (2443684) and the Sloan Research Fellowship.

\section*{Acknowledgement}
We thank Dr. Natasha Tomm and Dr. Philip Dolan for helpful input on the experimental method, and Prof. Andreas Muller for early test samples and helpful discussion on the cavity mounting method. We thank the JILA machine shop and Ryan Murdick for feedback on the assembly design. We thank the JILA Keck lab and NIST Boulder cleanroom staff for their help with sample processing.

\section*{Disclosures}
P.I., F.M., and S.S. are the Founder, Director of Operations, and Scientific Advisor, respectively, of Icarus Quantum Inc., which commercializes quantum-dot-based single-photon sources and quantum transducers. T.D.H., S.S., P.I., and F.M. are inventors on a pending patent application related to the open microcavity mechanical assembly described in this work.

\section*{Data availability}
Data underlying the results presented in this paper are not publicly available at this time but will be available via a DOI link upon the publication of this paper. 

\section*{Supplemental document}


\bibliography{bibliography}

\end{document}


\title{Supplementary Material: A Compact, Robust, and Tunable Open Microcavity Platform for Solid-State Quantum Electrodynamics}

\author{Thi D. Hoang\authormark{1}, Fateme Mahdikhany\authormark{2}, Zixuan Wang\authormark{3}, Richard Mirin\authormark{3}, Kevin Silverman\authormark{3}, Poolad Imany\authormark{2}, and Shuo Sun\authormark{1,*}}

\address{\authormark{1}JILA and Department of Physics, University of Colorado Boulder, Boulder CO 80309, USA\\
\authormark{2}Icarus Quantum Inc., Boulder 80302, USA\\
\authormark{3}National Institute of Standards and Technology, Boulder CO 80305, USA}

\email{\authormark{*}shuosun@colorado.edu}

\section{Epitaxial growth of the quantum dot substrate}
\begin{figure}[h!]
    \centering
    \includegraphics[width=0.6\linewidth]{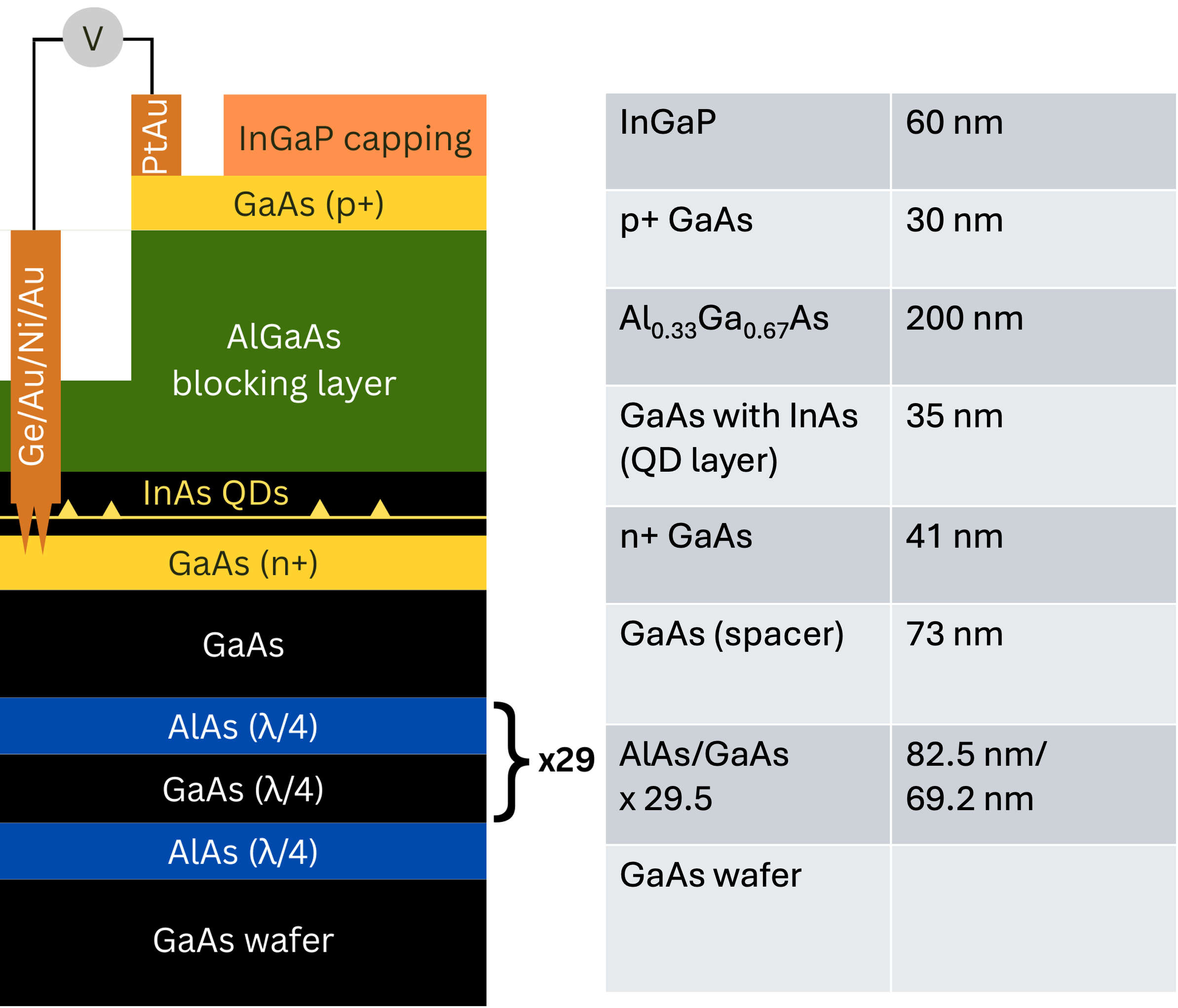}
    \caption{Schematics showing the layers of material on the quantum dot sample and their thicknesses.}
\label{fig:QDchip}
\end{figure}

The quantum dot substrate is grown by molecular beam epitaxy on a GaAs wafer, with the layer structure shown in Fig.~\ref{fig:QDchip}. We first grow 29.5 pairs of alternating AlAs and GaAs layers to form the bottom distributed Bragg reflector (DBR). Next, we grow the p-i-n junction with InAs quantum dots embedded in the intrinsic layer. The junction includes an \textit{n}-doped GaAs layer (Si-doped) with a thickness of 41~nm, followed by an intrinsic GaAs region that contains a monolayer of InAs quantum dots, located 30~nm above the \textit{n}-layer. We then grow a 200-nm-thick Al$_{0.33}$Ga$_{0.67}$As blocking layer above the quantum dots. Finally, we grow a \textit{p}-doped GaAs layer with a thickness of 30~nm. For this specific wafer, an additional 60-nm-thick carbon-doped
InGaP capping layer is grown on top.

After growth, electrical contacts to the \textit{n}- and \textit{p}-doped layers are fabricated for charge state control of the quantum dots. Ohmic electrical contacts to the \textit{n}-doped layer are created by first etching vias down to approximately 80 nm above the \textit{n}-layer, and then depositing 400-nm-thick Ge$/$Ni$/$Au$/$Ni via electron beam evaporation. Ohmic contacts to the \textit{p}-doped layer are created by depositing 200-nm-thick Pt$/$Au to the p-layer after etching down from the InGaP capping. Both metals are then annealed at 430 $^\circ$C in forming gas (5\% $\mathrm{H}_2$ in Ar) to ensure good ohmic contacts.

\section{Bias tuning of quantum dot transition frequency}

\begin{figure}[h!]
    \centering
    \includegraphics[width=0.8\linewidth]{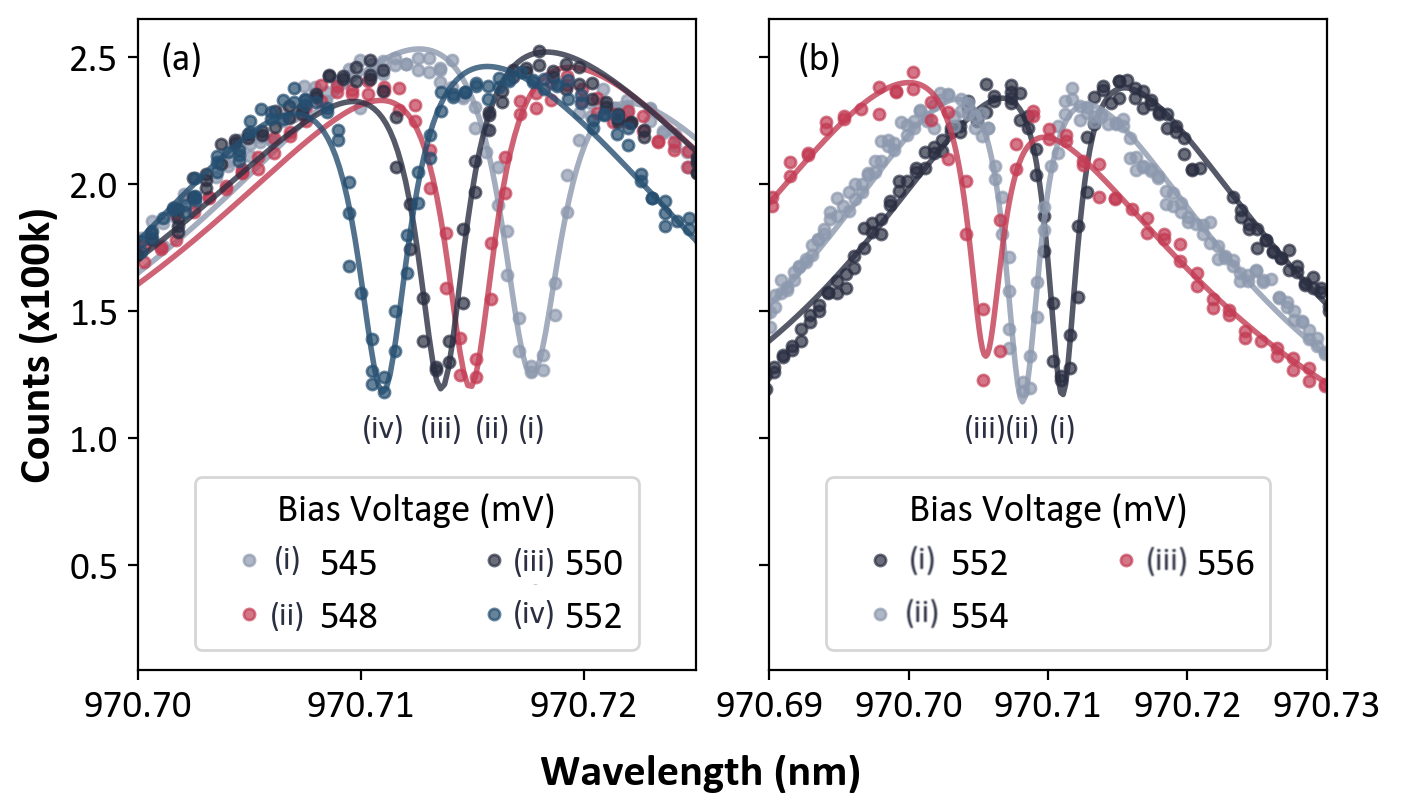}
    \caption{Cross-polarized transmission spectra of an open microcavity coupled to a quantum dot. (a) Spectra measured as the bias voltage applied across the p--i--n junction is varied, showing DC Stark tuning of the quantum-dot resonance. (b) Spectra measured while simultaneously tuning the quantum-dot resonance through the bias voltage and the cavity resonance through the piezo voltage, keeping the cavity resonant with the quantum dot at different wavelengths.}
    \label{fig:bias_tuning}
\end{figure}

The integration of the quantum dots with a p--i--n diode enables fine tuning of the quantum-dot resonance via the DC Stark effect by applying a bias voltage across the diode, and this capability is preserved for quantum dots coupled with our open microcavities. Fig.~\ref{fig:bias_tuning} shows cross-polarized transmission spectra of an open microcavity coupled to a quantum dot. The broad peak is the cavity response, whereas the narrow dip results from coupling to the quantum dot. Fig.~\ref{fig:bias_tuning}(a) shows the spectra as the bias voltage applied to the p--i--n junction is varied. We clearly observe a Stark shift of the quantum-dot-induced dip as the bias voltage is changed. Fig.~\ref{fig:bias_tuning}(b) shows the same device, but with the cavity resonance tuned at each bias voltage to remain resonant with the quantum dot, demonstrating simultaneous tuning of the cavity and quantum-dot resonance. This capability allows us to tune a resonantly coupled quantum-dot--cavity system to any wavelength within the DC-Stark tuning range of the quantum dot, which is typically on the order of 0.1~nm. 

\section{Fabrication of the concave mirrors}
\label{sec:mirror_fabrication}
Our concave mirrors are fabricated using the resist-reflow method~\cite{jin2022microfabricated}. On a fused silica substrate, we used optical lithography to define an array of circular disks of SPR660 photoresist ($1~\mu$m thick) on the substrate. The resist is then heated to 140$\degree$C for 30 minutes and becomes viscous, forming a concave shape in the middle of the original disk due to surface tension. For the specific mirror substrate employed in this work, we defined an array of $50 \times 50$ concave mirrors with a center-to-center spacing of $50~\mu$m. By varying the size of the initial resist disks across the array, we were able to control the radii of curvature (ROC) ranging from $50~\mu$m to $300~\mu$m. For results reported in the main text, we chose the mirrors with a radius of curvature around $65~\mu$m, as also shown in section~\ref{sec:mirror_surface}.

\begin{figure}[h!]
    \centering
    \includegraphics[width=0.7\linewidth]{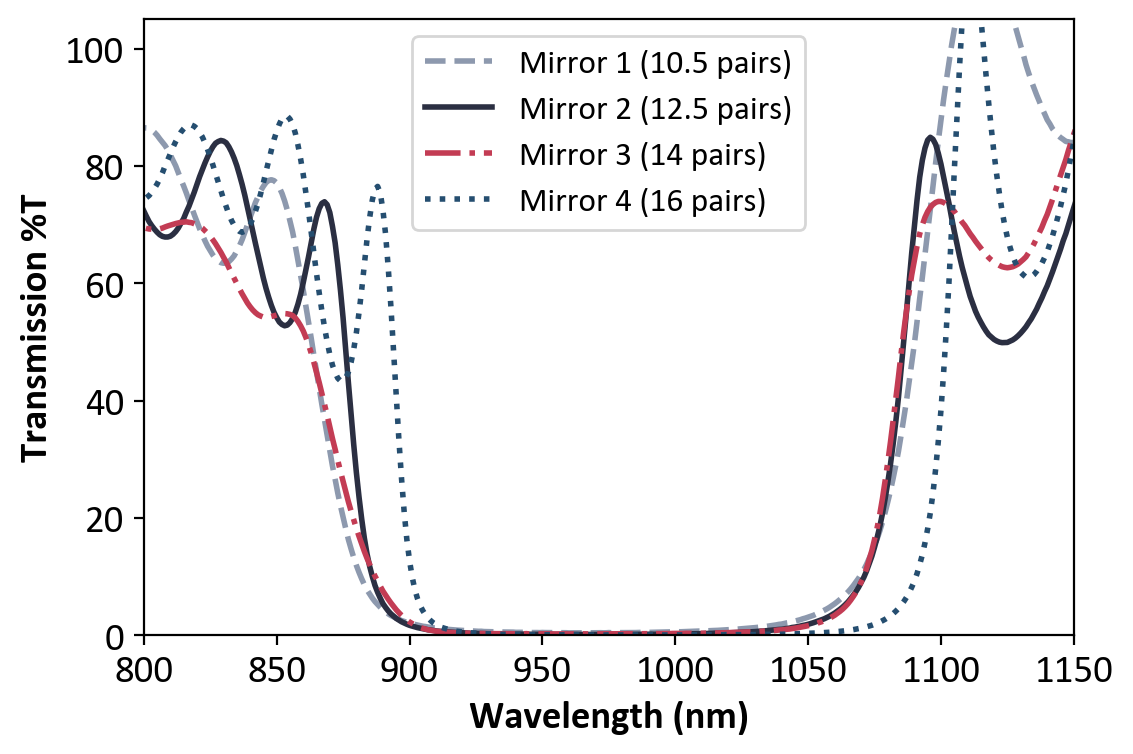}
    \caption{Spectrophotometer measurement of the transmission spectra for 4 different mirrors, each with a different number of pairs of SiO$_2$/SiN, coated with PECVD. The absolute transmission value was not calibrated, hence there is transmission exceeding 1. Mirror 2 (12.5 pairs) is used for experiments reported in the main text.}
    \label{fig:stopbands}
\end{figure}

After reflow, the resist patterns is then transferred onto the silica substrate through reactive ion etching, forming microscale concave dimples on the silica substrate. The whole substrate containing the microscale dimples is subsequently coated with 12.5 pairs of alternating $\text{SiO}_2$/SiN layers, each a quarter-wavelength thick ending with the higher-index material, by plasma-enhanced chemical vapor deposition (PECVD). We note that PECVD does not generally provide the same level of thickness control, layer-to-layer reproducibility, or surface quality as techniques such as ion-beam sputtering. To improve the reproducibility of the PECVD DBR coating, we took two calibration steps. First, before each DBR coating, we performed a test deposition on a dummy chip and calibrated the deposition rate and refractive index of both dielectric materials using ellipsometry. This allowed us to fine-tune the deposition parameters based on the condition of the PECVD system prior to the DBR coating. Second, after DBR coating, we measured the transmission stopband of the mirrors to verify that the stopband was close to the designed wavelength. Figure~\ref{fig:stopbands} shows the measured transmission spectra of PECVD DBRs from four different coating runs, with each run using a different number of mirror pairs. Qualitatively, the four coating runs yield similar stopband centers, confirming the consistency of the PECVD coating across runs.

\section{Surface characterization of concave mirrors}
\label{sec:mirror_surface}
\begin{figure}[h!]
    \centering
    \includegraphics[width=0.9\linewidth]{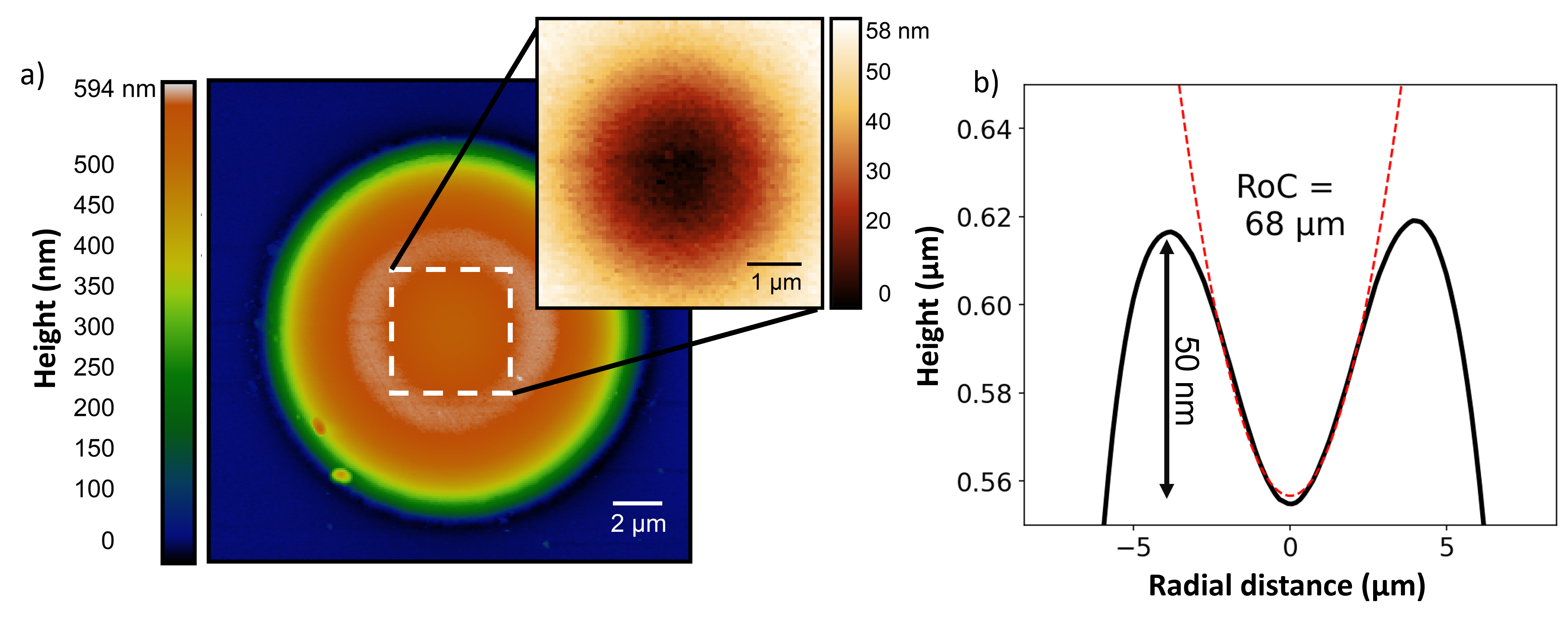}
    \caption{AFM and profilometer imaging of the concave feature. (a) AFM image showing the full profile of one reflow feature. Inset shows the central concave region of the reflow feature. The concave area measures $\approx 6~\mu$m across and $\approx 50$~nm deep. (b) Cross section from a stylus profilometer measurement of a similar device showing the concave profile at the center of the reflow feature. Fitted parabola shows a Radius of Curvature (RoC) of $68~\mu$m and a concave depth of 50~nm.}\label{fig:surface}
\end{figure}

Figure~\ref{fig:surface}(a) shows an atomic force microscope (AFM) image of the concave mirror fabricated via the method described in section \ref{sec:mirror_fabrication}. The inset shows the concave area used for the cavity, which has a depth of $\sim$50~nm (defined as the difference between minimum and maximum height in the area) and a diameter of $\sim$6~$\mu$m. Analysis at the concave area shows a rms area surface roughness of 0.5~nm, limited by the resolution of our AFM. This upper limit is an order of magnitude lower than the sub-\text{\AA} level quoted for ion sputtering coating~\cite{jin2022microfabricated}. As PECVD is known to produce coatings with higher roughness than ion sputtering, we expect the coating process to be the limiting factor for our surface roughness.

To quickly characterize the radius of curvature for different features, we used the stylus profilometer to scan across the concave features in one line. Figure~\ref{fig:surface}(b) shows one such scan at the center of a feature of the same size as in Fig.~S2(a). By fitting the center of the feature to a parabola, we extracted a radius of curvature of $68~\mu$m. The feature depth shown from the profilometer measurement is also consistent with the AFM measurement.

\section{Cavity assembling procedure}
\label{sec:assembling}

Once the top and bottom chips are set on their respective metal holders (e.g. glued with rubber cement), the mounting procedure starts with laying the top holder above the bottom holder and adding the corner washers and screws. The set screws that hold the top chip (through the metal aperture) are kept loose, letting the top chip freely tilt as it touches the bottom chip. This ensures a good initial parallelism between the two chips. To assist this further, we evenly apply a downward force onto the top ring as we screw in the set screws, fixing the top chip in place. We also used soft home-made Indium washers to facilitate fine tuning of the angle between the top and bottom mirrors during the assembly. If the chip surfaces are smooth and clean, interference fringes (so-called “Newton rings”) should immediately appear, visible to the naked eyes. 

\begin{figure}[h!]
    \centering
    \includegraphics[width=0.7\linewidth]{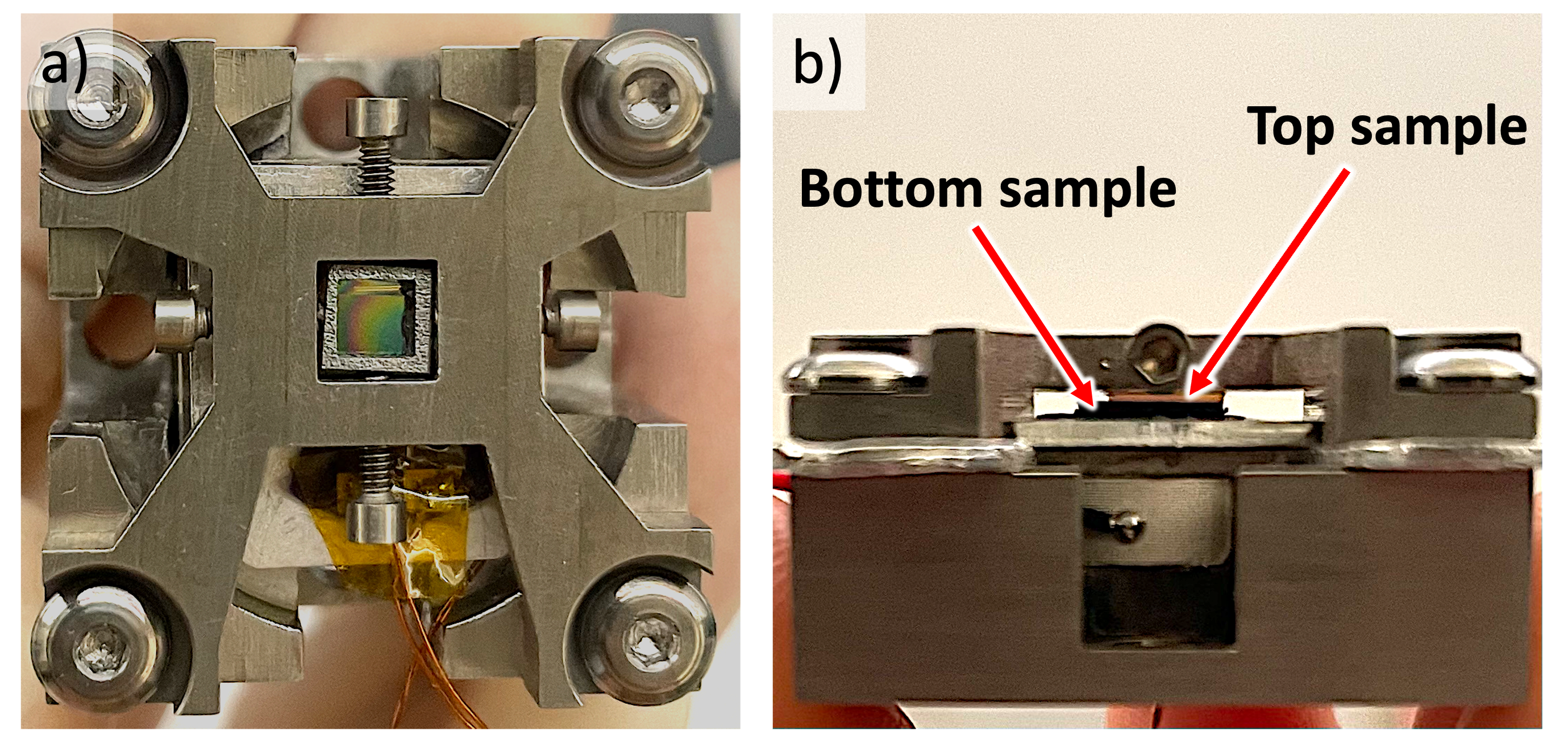}
    \caption{Mounting procedure for the open microcavity assembly. (a) Newton rings appear as two reflective samples are in contact. Image taken under white fluorescent room light. (b) Side view of the assembly, showing the two samples lying flat on top of each other. The two samples are the only two surfaces in contact.}
    \label{fig:mounting}
\end{figure}

Figure~\ref{fig:mounting} shows the Newton rings observed during the mounting procedure, as well as a side view of the assembly showing the two samples lying flat on top of each other. The two samples are the only two surfaces in contact, which is important for ensuring good parallelism and minimizing mechanical vibrations. The assembly design is such that the cavity length is the shortest at maximum piezo voltage. We keep the piezo fully extended during the above mounting procedure to ensure that the two mirrors are parallel at their shortest distance, so that the bottom mirror's downwards movement remains unobstructed when the piezo contracts, ensuring unobstructed tuning of the cavity length.

\section{Relation between cavity length and the measured free spectral range}
\label{sec:cavity-length}

For Fabry--Perot cavities formed with Bragg mirrors, the frequency-dependent reflection phase of each mirror can influence the cavity resonance condition, so the free spectral range is not determined solely by the physical mirror separation~\cite{koks2021microcavity}. This effect is particularly important for short cavities, where adjacent longitudinal modes can lie near opposite edges of the DBR stopband, where the mirror phase varies most strongly with frequency.

\begin{figure}[h!]
    \centering
    \includegraphics[width=0.8\linewidth]{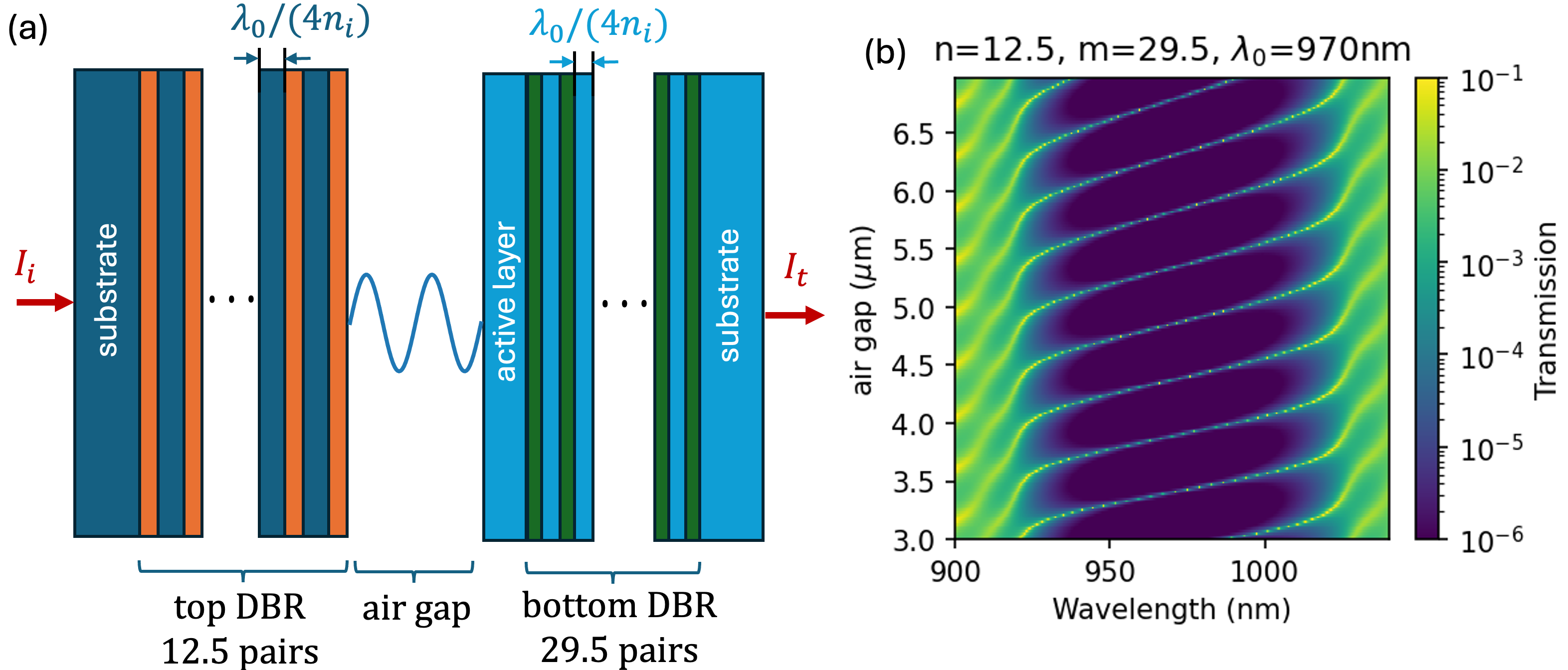}
    \caption{(a) Schematic of the cavity stack used in the experiment. The cavity stack consists of top and bottom DBRs made from different material systems (SiN/SiO$_2$ and GaAs/AlGaAs), with an air gap and a GaAs layer sandwiched between them. (b) Transfer matrix simulation of the cavity transmission, $T\equiv I_t/I_i$, for air-gap lengths ranging from 3 to 7~$\mu$m.}
    \label{fig:cavityL_sim}
\end{figure}

Using transfer matrix methods, we calculated the cavity resonances of our open microcavity, as illustrated in Fig.~\ref{fig:cavityL_sim}(a). The cavity consists of two DBRs made from different material systems (SiN/SiO$_2$ and GaAs/AlAs), with an air gap and an active GaAs layer between them. In the simulation, we assume that the two mirrors have the same stopband center at 970~nm. Figure~\ref{fig:cavityL_sim}(b) shows the simulated transmission spectra as we vary the air-gap length. The peak of each transmission spectrum, shown as the bright lines, indicates the resonant frequency of the cavity modes. Near the stopband center, the cavity resonance wavelength increases approximately linearly with increasing air-gap length. Near the stopband edges, however, the resonances deviate from this linear behavior because of mirror dispersion. This deviation leads to an apparent reduction in the extracted free spectral range, which in turn causes the cavity length inferred from the FSR to be overestimated. Since the cavity modes in Fig.~2(a) of the main text lie near the edges of the mirror stopband, we report the cavity length extracted from the FSR as an upper bound, as stated in Sec.~3 of the main text. In our simulations, this stopband-edge dispersion is the dominant correction, exceeding other effects such as the use of different mirror materials or slightly mismatched mirror stopbands.

\section{Cavity tuning range with alternative piezo actuator}
\begin{figure}[h!]
    \centering
    \includegraphics[width=0.7\linewidth]{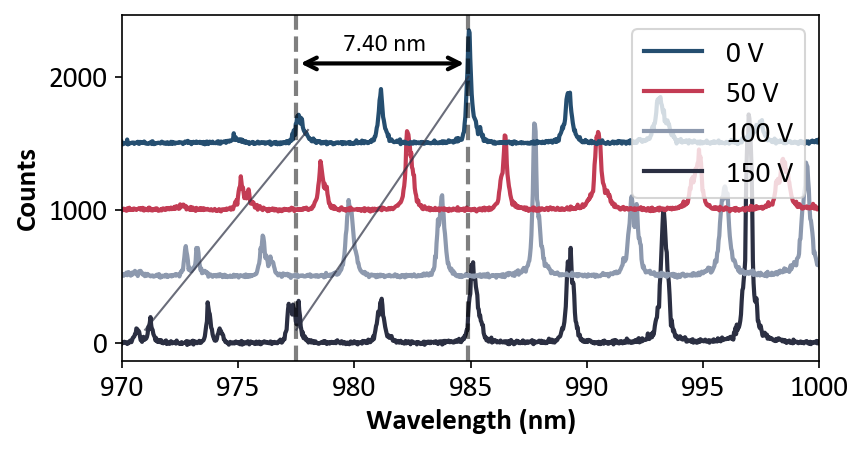}
    \caption{Spectra of the cavity resonances under photoluminescence at different piezo voltage, offset for visibility. The expanded tuning range compared to Fig. 2(d) in main text due to using a wider-range piezo actuator.}
    \label{fig:long_piezo}
\end{figure}

The modular design of the cavity allows for easy replacement of different parts of the design, including the piezo actuator used. We tested out the possibility of increasing the tuning range by replacing the Thorlabs ring piezo (PA44M3KW, free stroke room temperature range of $3.8~\mu$m) with a Noliac ring stack (NAC2123-H16, free stroke room temperature range of $23~\mu$m). With this modified assembly, we observed a cavity resonance tuning range of more than 7~nm, as seen in Fig.~\ref{fig:long_piezo}. The four spectra shows photoluminescence counts collected from cavity transverse modes ranging in wavelength from 970~nm to 1000~nm. 

For this measurement, we took the cavity spectrum at 4 different piezo voltages (0, 50 V, 100 V, 150 V). The diagonal line is a guide for the eyes to show the same cavity mode moving across different piezo voltages. Although only four spectra were recorded for this purpose, the peaks moved smoothly across piezo voltages during the experiment, allowing us to confidently identify the same mode across cavity lengths. The multimode spectrum is due to the presence of higher transverse modes in this cavity, which shows up due to the incoming beam being slightly misaligned from the cavity center to excite multiple modes.

Extracting the resonant wavelengths, we observed a tuning range of $7.6$~nm for one of the cavity modes, compared to $3.0$~nm for the shorter piezo used in our main text. Despite the longer tuning range, we did not use this assembly for the main experiment as the quantum dot chip used for this experiment was not of good quality.

\section{Resolving cavity mode splitting with photoluminescence spectroscopy}
\label{sec:mode-split-PL}

In contrast to the cavity transmission spectra shown in Figs.~2(b) and 2(c) of the main text, which were obtained by laser scanning, the cavity photoluminescence spectra shown in Figs.~2(a), 2(d), and 2(e) do not resolve the doublet arising from the two non-degenerate polarization modes. This difference is due to the finite spectral resolution of the spectrometer used for the photoluminescence measurements. Our spectrometer can reach a resolution of 0.03~nm when equipped with the highest-groove-density grating available to us, 1800~lines/mm. However, this grating has a cutoff wavelength of 950~nm and therefore cannot be used for the cavity reported in the main text, whose resonance wavelength is around 970~nm.

\begin{figure}[h!]
    \centering
    \includegraphics[width=0.5\linewidth]{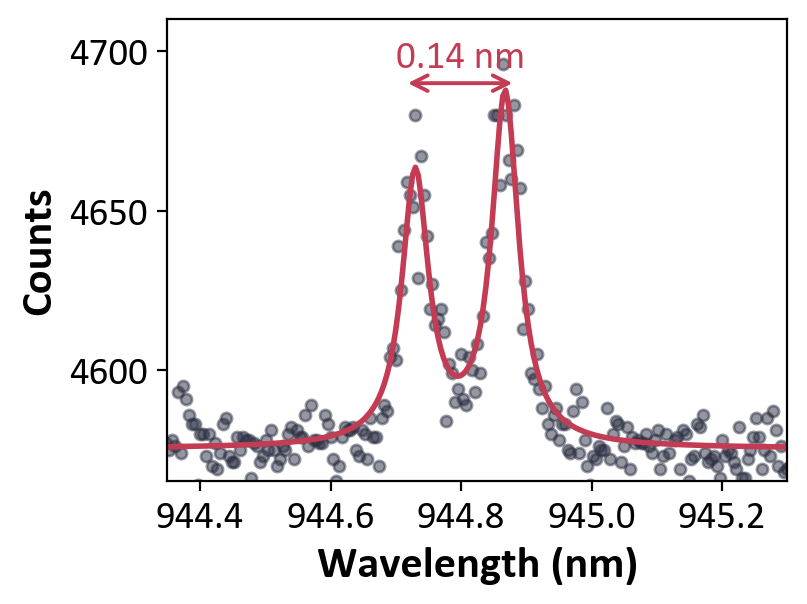}
    \caption{Photoluminescence spectrum of a different cavity device from the one shown in Fig.~2(b), showing the split cavity modes resolved on the spectrometer. The data were taken using a grating with a groove density of 1800~lines/mm.}
    \label{fig:PL_splitting}
\end{figure}

For completeness, Fig.~\ref{fig:PL_splitting} shows the photoluminescence spectrum of another cavity device with a resonance wavelength below 950~nm, where the high-resolution grating can be used. In this case, the photoluminescence spectrum clearly resolves the splitting between the two non-degenerate polarization modes.

\section{Theoretical modeling}
\label{sec:theory}

Figure~3(a) in the main text shows the cross-polarized cavity transmission spectrum for two cavity modes each coupling to a quantum dot. We model the cavity transmission as
\begin{equation}
    T(\omega_L) = A_H\mathrm{Tr}[\rho_{ss}(\omega_L)\op{a}^\dag \op{a}] + A_V\mathrm{Tr}[\rho_{ss}(\omega_L)\op{b}^\dag \op{b}] + A_0
    \label{eq:transmission}
\end{equation} 
where $\mathrm{Tr}[\rho_{ss}(\omega_L)\op{a}^\dag \op{a}]$ and $\mathrm{Tr}[\rho_{ss}(\omega_L)\op{b}^\dag \op{b}]$ represent the expected photon numbers in the non-degenerate cavity modes ($H-$ and $V-$polarized) with corresponding annhilation operators $\op{a}$ and $\op{b}$. Here, $\rho_{ss}(\omega_L)$ is the steady state density matrix of the system when both cavity modes are driven by a laser with frequency $\omega_L$. $A_H, A_V$ are scaling factors that relate the transmission intensity to the photon numbers, and $A_0$ is the overall intensity offset. $A_H$ and $A_V$ can be different due to different cross-polarization filtering efficiency and different output collection efficiency of the two polarizations.

To calculate the density matrix $\rho_{ss}$, we solve the Lindblad master equation of the system, given by
\begin{equation}
    \dot{\rho} = -i[H, \rho] + \sum_i \left(L_i \rho L_i^\dag - \frac{1}{2}\{L_i^\dag L_i, \rho\}\right)
\end{equation}
where $H$ is the system Hamiltonian, and $L_i$ are the jump operators for each loss channel. The system consists of two effective two-level systems (quantum dots) coupled to two non-degenerate cavity modes driven by a laser field. We ignore the coupling between each quantum dot to the off-resonant cavity mode due to large detunings. In the rotating frame of the laser, the Hamiltonian is
\begin{multline}
    H = (\omega_H-\omega_L)\op{a}^\dag \op{a}+ (\omega_V-\omega_L)\op{b}^\dag \op{b}+
    (\omega_1-\omega_L)\op{\sigma_1}\op{\sigma_1}^\dag
    + (\omega_2-\omega_L)\op{\sigma_2}\op{\sigma_2}^\dag\\
    + ig_H(\op{a}^\dag\op{\sigma_1}-\op{a}\op{\sigma_1}^\dag)
    + ig_V(\op{b}^\dag\op{\sigma_2}-\op{b}\op{\sigma_2}^\dag)
    +\eta_H(\op{a}^\dag+\op{a})+\eta_V(\op{b}^\dag+\op{b})
\end{multline}
where $\omega_H$ ($\omega_V$) is the resonant frequency of cavity mode $H$ ($V$) with annihilation operator $\op{a}$ ($\op{b}$), $\omega_1$ ($\omega_2$) is the dot transition frequency with lowering operator $\op{\sigma_1}$ ($\op{\sigma_2}$), $g_H$ ($g_V$) is the coupling strength between the cavity mode $H$ ($V$) and the resonant quantum dot 1 (2), and $\eta_H$ ($\eta_V$) is the coherent drive amplitude of the laser field on cavity mode $H$ ($V$). The open quantum system has the following jump operators, corresponding to the two cavity modes' dissipation, the spontaneous emission from the two quantum dots, and the pure dephasing of the two quantum dots, respectively:
\begin{gather*}
    L_a=\sqrt{\kappa_H}\op{a}, L_b=\sqrt{\kappa_V}\op{b}, \\
    L_1=\sqrt{\gamma_1}\op{\sigma_1}, L_2=\sqrt{\gamma_2}\op{\sigma_2}, \\
    L_{\text{d1}}=\sqrt{\gamma_{\text{d1}}}\op{\sigma_1}\op{\sigma_1}^\dag, L_{\text{d2}}=\sqrt{\gamma_{\text{d2}}}\op{\sigma_2}\op{\sigma_2}^\dag,
\end{gather*}
where $\kappa_H$ and $\kappa_V$ are the leakage rates for the two cavity modes $H$ and $V$, while $\gamma_1$, $\gamma_2$ are the spontaneous emission rates and $\gamma_{\text{d1}}, \gamma_{\text{d2}}$ are the pure dephasing rates of the two quantum dots. 

We fit the theoretical model to the spectrum in Fig.~3(a) using QuTiP and SciPy’s least-squares method. In the fit, we set $\kappa_H/2\pi=16.04$~GHz, $\kappa_V/2\pi=18.04$~GHz, $\omega_H/2\pi=309.0177$~THz, and $\omega_V/2\pi=309.0540$~THz, all extracted from the bare cavity spectrum shown in Fig.~2(b). We also set $\gamma_1/2\pi=\gamma_2/2\pi=0.16$~GHz, obtained from prior measurements of similar dots. We further assume that each quantum dot is resonant with the respective cavity mode, i.e. $\omega_1 = \omega_H$ and $\omega_2 = \omega_V$. Fig.~3(a) is obtained in the weak excitation limit ($\sim 1$~nW), so we keep $\eta_H$ and $\eta_V$ to be $\eta_H = \eta_V = 0.1$. We then fit the calculated spectrum from Eq.~\eqref{eq:transmission} to the full experimental spectrum with $g_1, g_2, \gamma_{\text{d1}}, \gamma_{\text{d2}}, A_H, A_V$, and $A_0$ as free parameters. The fitted values for coupling strengths and dephasing rates are $g_H/2\pi=1.37\pm0.02$~GHz, $g_V/2\pi=1.64\pm0.04$~GHz, $\gamma_{\text{d1}}/2\pi=0.05\pm0.04$~GHz, $\gamma_{\text{d2}}/2\pi=0.17\pm0.03$~GHz, corresponding to total decay rates $\Gamma_1 = \gamma_1 + 2\gamma_{\text{d1}} = 0.26\pm0.08$~GHz, $\Gamma_2 = \gamma_2 + 2\gamma_{\text{d2}} = 0.50\pm0.06$~GHz and cooperativities $C_H = 4g_H^2/(\kappa_H\Gamma_1) = 1.8\pm0.2$, $C_V = 4g_V^2/(\kappa_V\Gamma_2) = 1.2\pm0.1$.

The fitted model shows some discrepancy with the measured data, both in the spectral tails and in the dip contrast. For the dip contrast, the discrepancy likely arises because the model does not account for the quantum dot spectral diffusion. In the current model, the only free parameters that broaden the quantum dot are the pure dephasing rates $\gamma_{d1}$ and $\gamma_{d2}$, which describe homogeneous broadening. In reality, however, the quantum dot can be broadened by both homogeneous dephasing and inhomogeneous spectral diffusion. Since spectral diffusion is not included in the model, the model effectively assigns a larger homogeneous linewidth to reproduce the overall linewidth of the dip. This reduces the fitted dip contrast compared with the actual data, where the quantum dot may have a narrower homogeneous linewidth together with inhomogeneous spectral broadening.  

The discrepancy in the spectral tails most likely arises from a weak residual coherent background in the cross-polarized detection channel. In an ideal cross-polarization measurement, the directly reflected laser field is fully rejected, and the detected spectrum reflects only the cavity response. In practice, however, wavelength- and polarization-dependent phase shifts or losses in the optical path, cryostat windows, or local cavity surface can prevent perfect cancellation over the full spectral range. The detected field can therefore contain a coherent superposition of the desired cavity signal and a weak background field, leading to distortion of the cavity lineshape.

For all spectra in Fig.~3(b), we fix $\kappa_H, \gamma_1, \omega_H, \omega_1$,and $A_H$ to the same values as above for Fig.~3(a). In addition, we also fix $g_H/2\pi=1.39$~GHz, $\gamma_{\text{d1}}/2\pi=0.04$~GHz and perform fitting for $\eta_H$ and $A_0$. The measured data agree well with the theoretical model even though we only have these two fitting parameters.

For Fig.~3(c), fitting is done with the cavity resonant frequency $\omega_H$ as the only free parameter. All other parameters are fixed to values used in and obtained from the fitting to Fig.~3(a).

\section{Cavity coupling parameters across different devices}
\label{sec:more-DIT}

\begin{figure}[h!]
    \centering
    \includegraphics[width=0.9\linewidth]{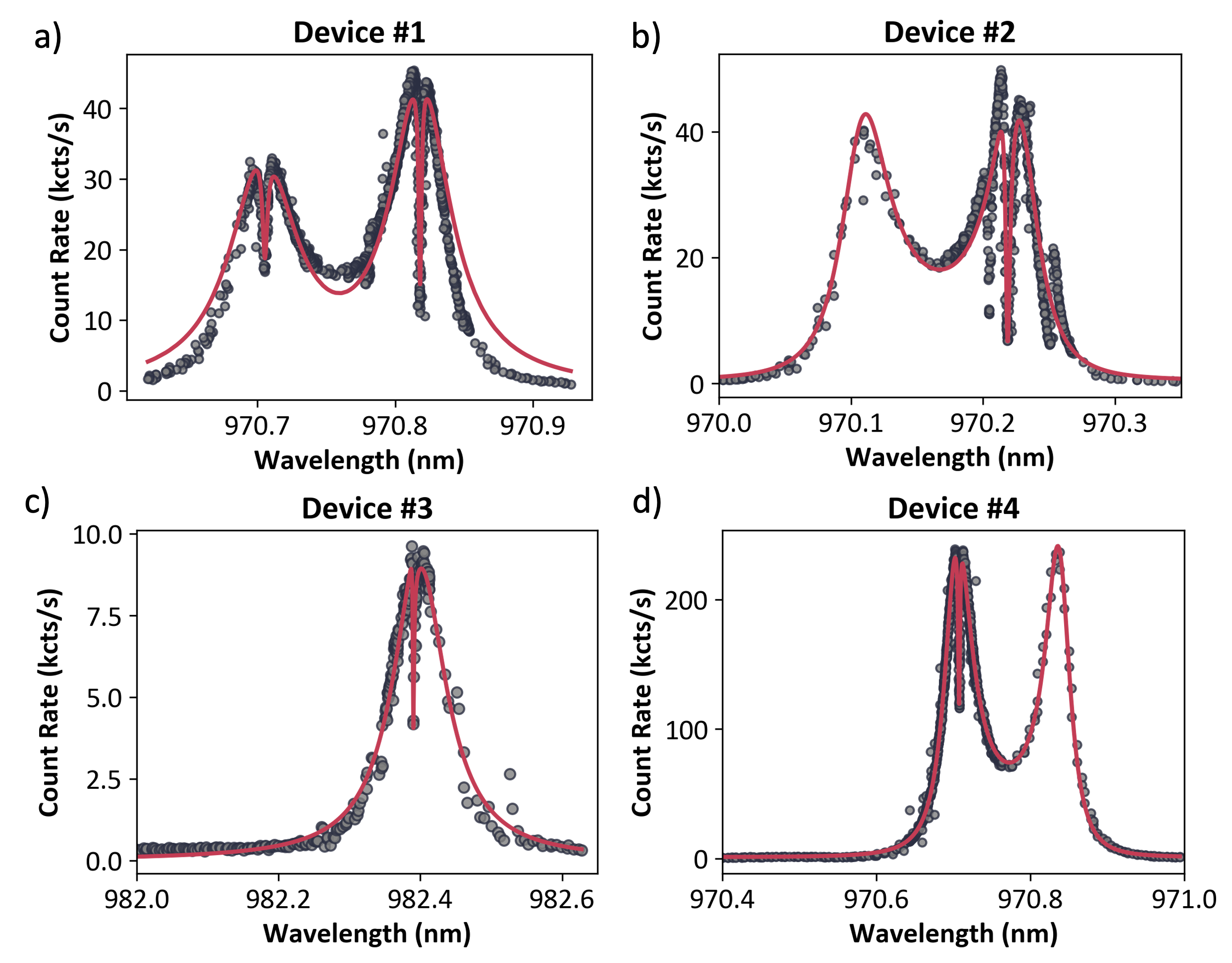}
    \caption{Cross-polarized transmission measurement of cavity QED devices with single quantum dots coupled to an open microcavity. Data on device \#1 also shown in Fig.~3(a) in main text. In all panels, gray dots are experimental data and red curves are theoretical fits.}
    \label{fig:moreDIT}
\end{figure}

Fig.~\ref{fig:moreDIT} shows the measured cavity transmission spectra for four devices, from which we extract the cavity coupling parameters summarized in Table~1 of the main text. Analysis for device \#1 is described in the main text and section~\ref{sec:theory}, recreated for comparison. Besides the fitted dip, spectra of devices \#2 and \#3 show additional dips and peaks, which arise from coupling between the cavity mode and other quantum dots not included in the numerical fitting model.

\section{Estimate of device yield per sample}
\begin{figure}[h!]
    \centering
    \includegraphics[width=\linewidth]{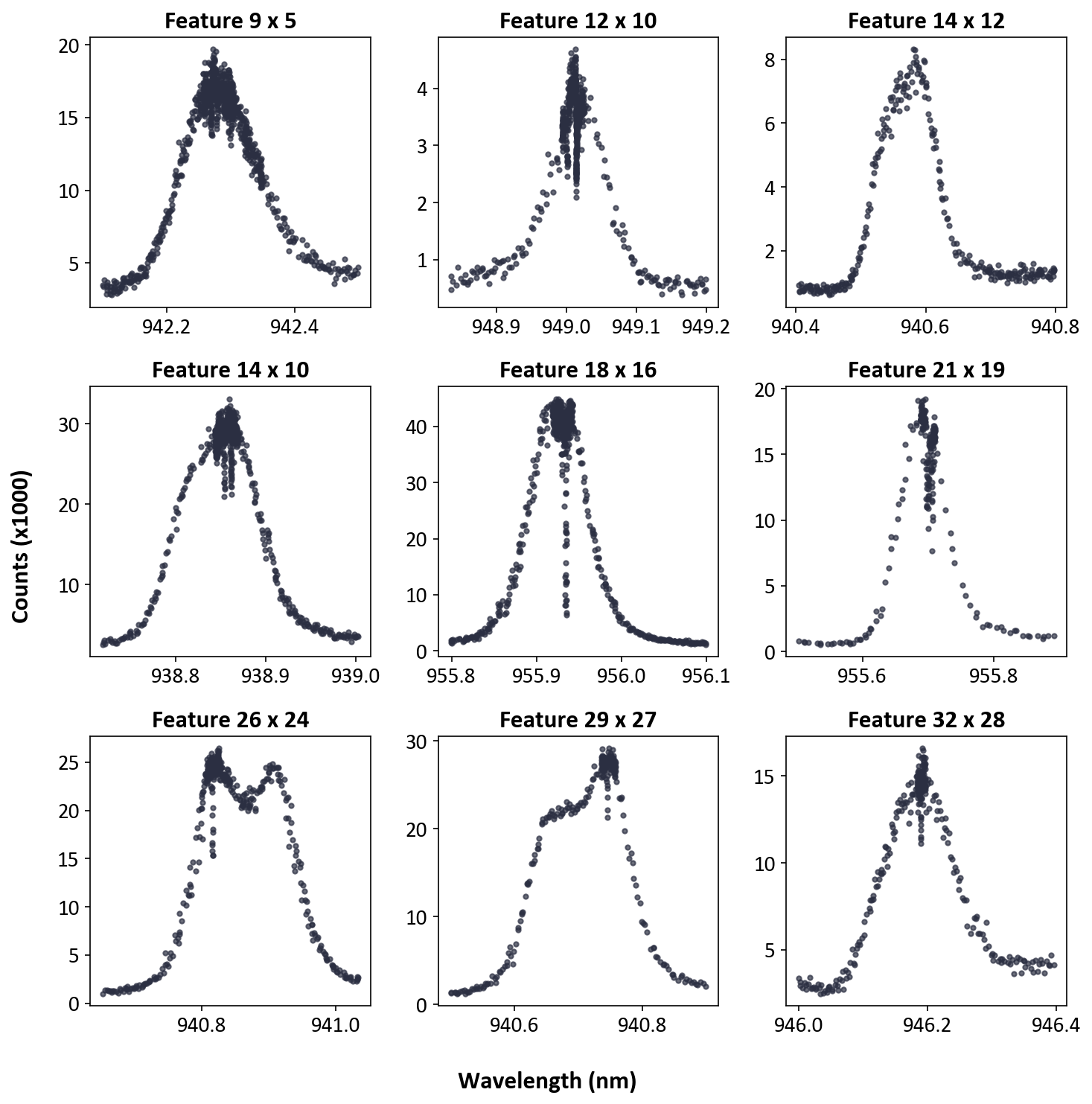}
    \caption{Cross-polarized cavity transmission spectra of 9 out of 10 randomly sampled cavities on the same assembly. The feature number indicates the position of the concave mirror in the array. A smaller row number corresponds to a smaller concave feature size and radius of curvature. The spectrum of feature 8x6 was not recorded because the cavity mode was overly broad and exhibited many strong higher-order modes. Excluding feature 8x6 and feature 14x12, 8 out of the 10 sampled features show at least one quantum-dot transition with a contrast higher than 25\%.}
    \label{fig:yield}
\end{figure}

To estimate the yield of devices that show clear emitter--cavity coupling, we randomly selected 10 cavities within the same assembly with radii of curvature ranging from $60~\mu$m to $300~\mu$m and resonances within the wavelength range of 930--970~nm. For each cavity, we used piezo tuning to identify quantum dots coupled to the cavity mode. Among the 10 devices, 1 exhibited many strong higher-order modes, suggesting a non-ideal concave mirror profile. The spectra of the remaining 9 cavities are shown in Fig.~\ref{fig:yield}. We found that 8 of these 9 cavities exhibited clear emitter--cavity coupling, with quantum-dot-induced dip contrasts of at least 25\%. The highest contrast among these devices is 80\%.

We note that these measurements were performed using the same top mirror chip but a different quantum-dot sample from the one reported in the main text, because the original quantum-dot chip was damaged. As shown in Fig.~\ref{fig:yield}, most cavity modes on this quantum-dot chip exhibit little or no mode splitting, in contrast to the clear non-degeneracy observed in Fig.~2(b) of the main text. The suppression of the cavity-mode splitting is likely due to the addition of a thicker strain-relief layer on top of the quantum-dot chip. We leave a detailed study of this effect to future work.

\label{sec:device-yield}

\bibliography{bibliography}